\documentstyle[pra,preprint,aps]{revtex}
\begin{document}
\draft
\title{Sterile Neutrinos and Supernova Nucleosynthesis}
\author{David O. Caldwell$^{1}$, George M. Fuller$^{2}$,
and Yong-Zhong Qian$^3$\footnote{On leave from T-5, MS B283,
Los Alamos National Laboratory, Los Alamos, NM 87545.}}
\address{
$^{1}$Institute for Nuclear and Particle Astrophysics and Cosmology, and
Department of Physics, University of California, Santa Barbara, CA 93106}
\address{
$^{2}$Department of Physics, University of California, San Diego,
La Jolla, CA 92093-0319}
\address{
$^{3}$School of Physics and Astronomy, University of Minnesota,
Minneapolis, MN 55455}
\date{\today}
\maketitle

\begin{abstract}
A light sterile neutrino species
has been introduced to explain simultaneously the
solar and atmospheric neutrino puzzles and the results of the LSND
experiment, while providing for a hot component of dark matter. Employing
this scheme of neutrino masses and mixings, 
we show how matter-enhanced active-sterile 
($\nu_{\mu,\tau} \rightleftharpoons \nu_s$) neutrino transformation
followed by active-active
($\nu_{\mu,\tau} \rightleftharpoons \nu_e$) neutrino transformation can
solve robustly the neutron deficit problem encountered by
models of
$r$-process nucleosynthesis associated with neutrino-heated supernova ejecta.
\end{abstract}
\pacs{PACS number(s): 14.60.Pq, 14.60.St, 26.30.+k, 97.60.Bw}
\vfil\eject

\section{Introduction}

In this paper we show how the invocation of sterile neutrinos in a novel
transformation scenario can enable the production of the heavy 
rapid-neutron-capture
($r$-process \cite{bbfh57}) elements in neutrino-heated ejecta from supernovae.
Interestingly, the current hints of (and constraints on) neutrino 
oscillation phenomena
from considerations of solar \cite{solar} and atmospheric 
neutrinos \cite{atmos} 
and from the
LSND experiment \cite{lsnd} are difficult to explain with only two 
independent neutrino
mass-squared differences (corresponding to three active neutrino 
species) \cite{fit}. By
contrast, the data are explained readily in terms of neutrino 
oscillations with three
independent mass-squared differences, corresponding to four neutrino 
masses \cite{4fit,barg}.
However, the observed width of the Z$^0$ \cite{width} can accomodate 
only three light,
active neutrinos. Therefore, a fit to the data requires 
introducing at least one light
\lq\lq sterile\rq\rq\ neutrino species which does not have normal 
weak interactions
[e.g., a Majorana SU(2) singlet neutrino].

The most promising site for $r$-process nucleosynthesis is the neutrino-heated
material ejected relatively long ($\sim 10\,{\rm s}$) after the 
explosion of a Type II
or Type Ib/c supernova \cite{nuheat}. However, detailed calculations 
of the conditions
which obtain above the neutron star remnant in such $r$-process
models show that 
the neutron-to-seed
ratio, $R$, is too low to allow the production of the heaviest 
$r$-process species
\cite{seed}. (The \lq\lq seed\rq\rq\ nuclei which capture neutrons to 
make the heavier
species have nuclear mass numbers between $50$ and $100$.) We require 
$R > 100$ to effect
a good $r$-process yield for heavy nuclear species, but the models 
with conventional
neutrino physics and conventional equations of state for nuclear matter
all give smaller
values of $R$. This is the neutron deficit problem for these models
of the $r$-process. 

In all models, $R$ is determined by the net 
electron-to-baryon number (i.e., the electron fraction),
$Y_e$, the entropy-per-baryon in the ejecta, $S$, and the 
dynamic expansion 
timescale, $\tau_{\rm
DYN}$, associated with the material ejection process \cite{seed}.
Though general relativistic effects \cite{gr} and
multi-dimensional hydrodynamic outflow \cite{multid} have been 
invoked to increase $S$
and decrease $\tau_{\rm DYN}$ (both of these changes favoring larger 
$R$) enough to solve
the neutron deficit problem, these solutions are at best finely 
tuned. On general grounds
we could argue that the only robust way to obtain $R>100$ in 
conventional {\it
neutrino-heated} ejecta is to decrease $Y_e$ 
and/or maintain its low value by
invoking new neutrino physics, e.g., introducing a sterile neutrino
as discussed in this paper. (By 
\lq\lq robust\rq\rq\ we mean
robust to astrophysical uncertainties in the detailed characteristics of
neutrino-heated outflow.) We cannot definitively proclaim at this 
point that new neutrino
physics is required to understand the production of heavy $r$-process elements. 
Such a proclamation calls for the accomplishment of 
the following. First, we would
have to establish that at least some of the $r$-process material originates in
conventional neutrino-heated ejecta. Second, we would have to 
understand the thermal
and hydrodynamic evolution of the very late stage 
neutrino-driven ``winds'' from a proto-neutron star.

The proposition that some of the $r$-process material
comes from environments with 
intense neutrino
fluxes is supported by recent studies of neutrino effects 
during and immediately following the $r$-process \cite{mmf98,nuproc}. 
Perhaps more to the point, recent
observational data on the abundances of
$r$-process species in old, very metal-poor halo stars in the Galaxy 
seem to be quite
consistent with what is expected from the $r$-process scenario
associated with neutrino-heated ejecta
\cite{sneden}. However, the second issue we would have to 
resolve is a vexing one. Despite the extensive numerical insights
into the early phase of the supernova evolution \cite{SN}, 
consistent and accurate hydrodynamic scenarios for neutron 
stars to lose mass through
neutrino heating at very late times remain to be established. 
On the positive side, Burrows
\cite{burrows} has recently followed his supernova calculations long 
enough to see the
formation of a neutrino-driven
wind at late times. In the region relevant for
nucleosynthesis, the wind in his calculations
resembles in broad brush the outflow models studied here.

This paper is organized as follows. In Sec. II, we describe a simple model
for the neutrino-heated outflow and discuss the associated alpha effect
which causes the neutron deficit problem for $r$-process nucleosynthesis.
In Sec. III, we discuss the treatment of active-sterile plus active-active
neutrino transformation in supernovae and describe how such transformation
can evade the alpha effect. The consequence of such transformation for
the evolution of the electron fraction $Y_e$
in the neutrino-heated outflow is
studied in Sec. IV. Conclusions are given in Sec. V.

\section{Neutrino-Heated Outflow and the 
Alpha Effect}

In what follows we adopt a simple exponential wind model for the
neutrino-driven outflow above the surface of the hot proto-neutron 
star produced by a
supernova explosion. In this model the enthalpy per baryon is roughly 
the gravitational
binding energy of a baryon, leading to a relation between radius 
$r_6$ (in units of
${10}^6\,{\rm cm}$), entropy-per-baryon $S_{100}$ (in units of $100$ 
times the Boltzmann
constant), and temperature $T_9$ (in units of ${10}^9\,{\rm K}$):
\begin{equation}
\label{eqn:wind1}
r_6 \approx {{22.5}\over{T_9 S_{100}}} \left({{M_{\rm NS}}\over{1.4 
M_\odot}}\right),
\end{equation}
where $M_{\rm NS}$ is the mass of the proto-neutron star.
Therefore, for a constant entropy per baryon characterizing the 
adiabatic expansion of the outflow,
the temperature parameterizes the radius \cite{gr}. The 
entropy typically
will be carried almost exclusively by relativistic particles (photons and
electron-positron pairs) so that $S\approx (2\pi^2/45) g_s T^3/(\rho 
N_A)$, where $g_s$
is the statistical weight for relativistic particles and
$N_A$ is Avogadro's number. This leads to a 
simple relation
between (matter) density $\rho$ and radius:
\begin{equation}
\label{eqn:dens}
\rho \approx 3.34\times{10}^3\,{\rm g\,cm^{-3}}
{\left( {{g_s}\over{11/2}}\right)} T_9^3 S_{100}^{-1} \approx 
3.8\times{10}^7\,{\rm
g\,{cm^{-3}}} {\left({{M_{\rm NS}}\over{1.4 M_\odot}}\right)}^3 {\left(
{{g_s}\over{11/2}}\right)} S_{100}^{-4} r_6^{-3},
\end{equation}
where we have scaled $g_s$ assuming its value at $T_9
{\ \lower-1.2pt\vbox{\hbox{\rlap{$>$}\lower5pt\vbox{\hbox{$\sim$}}}}\
}10$. As we will be
mostly interested in processes occurring at $T_9
{\ \lower-1.2pt\vbox{\hbox{\rlap{$>$}\lower5pt\vbox{\hbox{$\sim$}}}}\
}10$, the dependence
on $g_s$ will be suppressed hereafter.

In the exponential wind the radius of an outflowing mass element
is related to time $t$ by $r=r_0 
\exp{[{{\left(t-t_0\right)}/\tau_{\rm DYN}}]}$. Here
$\tau_{\rm DYN}$ is an assumed constant material expansion timescale. 
This implies an
outflow velocity proportional to radius, $v=r/\tau_{\rm DYN}$, so that the
neutrino-driven wind will remain roughly self similar for timescales 
on which the
neutrino luminosities and energy spectra and the neutron star radius can 
be regarded as
constant. Clearly, at some point above the neutron star 
surface the exponential
wind must go over to a linear expansion of the radius, i.e.,
a \lq\lq coasting\rq\rq\ outflow. Nevertheless, the 
exponential wind regime should
encompass the region above the neutron star where most of the biggest 
obstacle to
successful $r$-process nucleosynthesis, the alpha effect (see below), 
is operative
\cite{mmf98}.

In neutrino-heated ejecta, neutrino interactions with matter
supply the requisite energy for ejection of nucleosynthesis products.
The total amount of heating through these interactions determines
$S$ and $\tau_{\rm DYN}$ in the ejecta.
Most of this heating occurs
close to the
neutron star and $S$ and $\tau_{\rm DYN}$ are set at $T_9\sim 20$.
However, the dominant interactions for neutrino heating
\begin{equation}
\label{eqn:n}
\nu_e + n \to p +e^-
\end{equation}
and
\begin{equation}
\label{eqn:p}
\bar\nu_e + p \to n +e^+
\end{equation}
have prolonged effects on the neutron-to-proton ratio
($n/p=1/Y_e-1$) in the ejecta. First of all,
in the region where $T_9
{\ \lower-1.2pt\vbox{\hbox{\rlap{$>$}\lower5pt\vbox{\hbox{$\sim$}}}}\
}10$ and free nucleons are favored by nuclear statistical equilibrium
(NSE), the competition between the processes in Eqs. (\ref{eqn:n}) and
(\ref{eqn:p}) leads to
$n/p\approx \lambda_{\bar\nu_e p}
/\lambda_{\nu_e n}
\sim \left( L_{\bar\nu_e} \langle E_{\bar\nu_e} 
\rangle\right)/\left( L_{\nu_e}
\langle E_{\nu_e} \rangle\right)$. Here $\lambda_{\bar\nu_e p}$ and
$\lambda_{\nu_e n}$ are the rates for the reactions in Eqs.\ 
(\ref{eqn:p}) and
(\ref{eqn:n}), respectively, $L_{\bar\nu_e}$ and
$L_{\nu_e}$ are the $\bar\nu_e$ and $\nu_e$
energy luminosities, respectively, and
$\langle E_{\bar\nu_e} \rangle$ and 
$\langle E_{\nu_e}
\rangle$ are the average energies 
characterizing the corresponding neutrino energy spectra. 
Absent neutrino oscillations
(flavor/type mixings),
we expect $\langle E_{\nu_{\mu}}\rangle \approx \langle 
E_{\bar\nu_{\mu}}\rangle \approx
\langle E_{\nu_{\tau}}\rangle \approx \langle E_{\bar\nu_{\tau}} 
\rangle >\langle
E_{\bar\nu_e} \rangle >\langle E_{\nu_e} \rangle$ and, hence,
$n/p>1$ ($Y_e<0.5$, corresponding to neutron-rich ejecta)
\cite{qfwmw93}. We note that the numerical value of the $n/p$ ratio
depends on careful calculation of the rates
$\lambda_{\bar\nu_e p}$ and
$\lambda_{\nu_e n}$ \cite{horowitz} and accurate determination
of $L_{\bar\nu_e}$, $L_{\nu_e}$, $\langle E_{\bar\nu_e} \rangle$, and
$\langle E_{\nu_e}
\rangle$ \cite{transport}. However, the alpha effect discussed below
is insensitive to the value of the $n/p$ ratio obtained at $T_9
{\ \lower-1.2pt\vbox{\hbox{\rlap{$>$}\lower5pt\vbox{\hbox{$\sim$}}}}\
}10$.

The equilibrium of the $n/p$ ratio with
the $\nu_e$ and $\bar\nu_e$ fluxes 
described above is maintained until the ejecta
passes the weak freeze-out radius (at $T_9\sim 10$), $r_{\rm WFO}$, 
beyond which 
$\lambda_{\nu_e n}$,
$\lambda_{\bar\nu_e p}<1/{\tau_{\rm DYN}}$ and free nucleons
are no longer favored by NSE \cite{qfwmw93}.
However, even beyond $r_{\rm WFO}$, $\nu_e$ capture on neutrons
can force down the $n/p$ ratio and, hence, the neutron-to-seed ratio
$R$. This is the so-called \lq\lq
alpha effect\rq\rq\ \cite{fm95,mmf98}: as material flows to 
regions where the composition favored by NSE
shifts from free nucleons to a mixture of 
free nucleons and alpha
particles, protons will be incorporated into alpha particles, 
leaving
a disproportionate fraction of neutrons exposed to the intense flux 
of $\nu_e$. The 
process in Eq.\ (\ref{eqn:n}) then converts some of these neutrons into 
protons (which are
then immediately incorporated into alpha particles), thus progressively 
lowering the $n/p$ ratio and,
ultimately, $R$. 

The cumulative damage done to the $n/p$ ratio by $\nu_e$ 
capture on neutrons beyond the weak freeze-out radius $r_{\rm WFO}$ can
be estimated crudely by integrating $dY_e/dt \approx \lambda_{\nu_e 
n}(t) (1-2 Y_e)$
to obtain
\begin{equation}
\label{eqn:yet}
Y_e(t) \approx 0.5 + [ Y_e(0) -0.5]\exp{\left\{ -\lambda_{\nu_e n}(0)
\tau_{\rm DYN}[1-\exp(-2t/\tau_{\rm DYN})]\right\}},
\end{equation}
where initial values for the $\nu_e$ capture rate on neutrons and the electron
fraction, $\lambda_{\nu_e n}(0)$ and $Y_e(0)$, respectively, can be 
approximated as their values
at $r_{\rm WFO}$. Equilibrium of the $n/p$ ratio with
the $\nu_e$ and $\bar\nu_e$ neutrino 
fluxes
at early times requires that $\lambda_{\nu_e 
n}(0)
\tau_{\rm DYN} \sim 1$.  
Furthermore, the product
$\lambda_{\nu_e n}(0)
\tau_{\rm DYN}$ should be roughly constant for all models with
conventional neutrino
physics, since increased (decreased) neutrino luminosity will tend to
increase (decrease)
$\lambda_{\nu_e n}$ and decrease (increase) $\tau_{\rm DYN}$
proportionately.
[Neutrino heating through the processes in Eqs.\ (\ref{eqn:n}) and
(\ref{eqn:p}), which set $Y_e$, is
the principal determinant of $\tau_{\rm DYN}$.]
Therefore, from Eq.\ (\ref{eqn:yet}) it can 
be seen that
regardless of the freeze-out value 
$Y_e(0)$, sufficient exposure to the $\nu_e$ flux above $r_{\rm WFO}$
will drive $Y_e$
close to 0.5 ($n/p=1$,
i.e., no neutron excess). 

In summary, a great paradox exists for $r$-process nucleosynthesis in a
neutrino-driven wind. In such a scenario the neutrinos must supply 
enough energy [largely
through the processes in Eqs.\ (\ref{eqn:n}) and (\ref{eqn:p})] 
to lift baryons out
of the gravitational
potential well of the neutron star. The gravitational binding 
energy per nucleon
near the neutron star surface is $\sim 100\,{\rm MeV}$. Since the 
average $\nu_e$ and
$\bar\nu_e$ energies are $\sim 10\,{\rm MeV}$, a nucleon must 
interact with neutrinos
$\sim 10$ times in order to acquire enough energy for ejection from
the neutron star. In
turn, this requires a sufficiently large value of
$\lambda_{\nu_e n}(0)\tau_{\rm DYN}$, thus providing
conditions for a pernicious alpha effect which will cause a neutron
deficit. Such a neutron deficit will preclude a successful
$r$-process, especially the
synthesis of the heaviest nuclear species.

However, it is obvious that removal of the $\nu_e$ flux could neatly
solve the neutron deficit problem by unbalancing the competition 
between the 
processes in Eqs.\ (\ref{eqn:n}) and (\ref{eqn:p}) in favor of 
neutron production (thus lowering $Y_e$) and by disabling
the alpha effect in regions beyond $r_{\rm WFO}$ (thus maintaining
a low $Y_e$). Furthermore, as neutrino heating is essentially completed
at $T_9\sim 20$ (quite close to the neutron star surface),
the beneficial effects of
removing the $\nu_e$ flux
can be obtained without affecting the general thermal
and hydrodynamic characteristics of the neutrino-heated outflow 
(see Sec. III).

\section{Evasion of the Alpha Effect via 
Active-Sterile plus Active-Active Neutrino Transformation}

The existence of at least one light sterile neutrino could provide
a means to reduce the $\nu_e$ flux at sufficiently large radius to 
leave the process of
neutrino-heated ejection unhindered, yet sufficiently near or inside 
$r_{\rm WFO}$ so as
to disable the alpha effect, and thereby
fix the neutron deficit problem for $r$-process nucleosynthesis.
This $r$-process solution can be obtained by having an active-sterile
($\nu_{\mu,\tau}\rightleftharpoons \nu_s$) neutrino mass-level crossing
followed by 
an active-active ($\nu_{\mu,\tau}\rightleftharpoons \nu_e$)
neutrino mass-level crossing. 
(Hereafter, $\nu_s$ and
$\bar\nu_s$ denote left- and right-handed Majorana
sterile neutrinos, respectively.)
The first mass-level crossing converts $\nu_\mu$ and $\nu_\tau$,
which are emitted with the highest average energy from the neutron star,
into harmless sterile neutrinos. Without this, significant conversion of
$\nu_\mu$ and $\nu_\tau$ into $\nu_e$ would drive the 
neutrino-heated supernova ejecta proton rich \cite{qfwmw93}. 
With the supernova
$\nu_\mu$ and $\nu_\tau$
``sterilized'' by the first mass-level crossing, 
the second mass-level crossing now only
acts to convert $\nu_e$ emitted from the neutron star into 
$\nu_\mu$ and $\nu_\tau$. Charged-current capture reactions on neutrons
are energetically forbidden for supernova $\nu_\mu$ and $\nu_\tau$.
The net result is that the $\nu_e$ flux is reduced or removed, and
the alpha effect cannot operate. Therefore, the neutron excess is
preserved or possibly enhanced in this scenario.

The sequence of neutrino mass-level
crossings described above
can occur in a four-neutrino scheme \cite{4fit} which has 
two nearly degenerate
neutrino doublets separated by a mass-squared difference chosen to be 
compatible with the
LSND $\nu_\mu \rightleftharpoons \nu_e$ signal. In this scheme the 
lower mass neutrino pair would
either give an active-sterile $\nu_e
\rightleftharpoons \nu_s$ mass-level crossing in the sun, or  
provide a \lq\lq just so\rq\rq\ 
vacuum mixing
solution with $\delta m_{es}^2\sim 10^{-10}$~eV$^2$
to the solar neutrino problem \cite{BKS}. 
For solving the solar neutrino problem via
matter-enhanced $\nu_e\rightleftharpoons \nu_s$ mixing,
the mass-squared splitting for the lower mass neutrino pair must be 
$\delta
m^2_{es} {\ 
\lower-1.2pt\vbox{\hbox{\rlap{$<$}\lower5pt\vbox{\hbox{$\sim$}}}}\ }
{10}^{-5}\,{\rm eV}^2$ \cite{BKS}. The mass-squared splitting for the
higher mass neutrino pair is 
chosen to explain the
observed atmospheric $\nu_\mu/\nu_e$ data via maximal $\nu_\mu 
\rightleftharpoons \nu_\tau$
vacuum mixing.

Figure 1 shows our adopted neutrino mass scheme. 
The mass-squared splittings shown
are chosen to meet all experimental constraints and to solve the neutron
deficit problem for
$r$-process nucleosynthesis associated with 
neutrino-heated supernova ejecta. We would require 
$m_{\nu_s}  >
m_{\nu_e}$ for matter-enhanced
$\nu_e\rightleftharpoons \nu_s$ mixing in the sun, but a \lq\lq just 
so\rq\rq\ solution to the solar neutrino problem
could have $m_{\nu_s} < m_{\nu_e}$ as well. (Issues of compatibility 
with Big Bang
Nucleosynthesis limits aside, note that some kinds of singlet
\lq\lq sterile neutrinos\rq\rq\ could possibly evade bounds on the 
``just so'' $\nu_e\rightleftharpoons \nu_s$ vacuum 
mixing solution stemming from the SuperK and the 
Chlorine Experiments
\cite{mcng}.) In our overall mass scheme 
$\nu_\mu$ and $\nu_\tau$ could share the role of providing 
a hot dark matter
component \cite{Primack}.

Of the possible four-neutrino mass patterns, this one is most successful in
evading constraints. It has been shown that limits from accelerator and reactor
experiments disfavor having one dominant neutrino mass, 
e.g., a $3+1$ or a $1+3$
arrangement \cite{BGG}. A two-doublet scheme with $\nu_\mu 
\rightleftharpoons
\nu_s$ mixing (the higher mass pair)
explaining the atmospheric neutrino anomaly and $\nu_e 
\rightleftharpoons
\nu_\tau$ mixing (the lower mass pair) explaining the solar neutrino puzzle
is in trouble with Big Bang
Nucleosynthesis bounds \cite{Shi}.

With an appropriate choice of mixing angles and mass-squared difference
for the splitting of the doublets, the neutrino mass scheme in Fig. 1 
will lead to an
efficient matter-enhanced $\nu_{\mu,\tau} 
\rightleftharpoons
\nu_s$ transition above the neutron star surface 
(and the neutrino sphere), yet 
below the region where
an ordinary Mikheyev-Smirnov-Wolfenstein (MSW) \cite{msw} 
matter-enhanced $\nu_{\mu,\tau}\rightleftharpoons\nu_e$ mass-level
crossing would occur \cite{pelt}. 
Furthermore, the mass-level crossing for $\nu_e$
with energies most relevant for determining $Y_e$ can still
lie in the region
near or below the weak freeze-out radius $r_{\rm WFO}$.

\subsection{General Description of Neutrino Flavor/Type 
Evolution in Supernovae}

As discussed in Sec. IIIB, the overall problem of active-sterile 
plus active-active neutrino transformation in our case can be treated as
that of two separate mass-level crossings, 
with each mass-level crossing involving
effectively only two neutrino flavors/types.
The neutrino flavor/type evolution 
through each mass-level crossing is governed
by a Schroedinger-like equation: 
\begin{equation}
\label{eqn:schr}
i {d \over dt}{ \left [\matrix{
a_{x} (t)\cr
a_{y}(t)\cr
}\right ]  } = \left( H_{\rm v}
+H_{e} + H_{\nu \nu}\right) \left [\matrix{
a_x (t)\cr
a_y (t)\cr
}\right ],
\end{equation}
where $a_x(t)$ and $a_y(t)$ are the time-dependent amplitudes
for the neutrino to be in,
for example, the flavor eigenstates $|\nu_e\rangle$ and $|\nu_\mu\rangle$,
respectively.
The propagation 
Hamiltonian in Eq. (\ref{eqn:schr})
is given as the sum of three terms $H_{\rm v}$, $H_e$, and $H_{\nu\nu}$
resulting from vacuum masses, forward scattering on electrons and 
nucleons, and
forward scattering on \lq\lq background\rq\rq\ 
neutrinos, respectively.

In general, we can express the
neutrino propagation Hamiltonian as:
\begin{equation}
\label{eqn:H}
H_{\rm v}+H_{e} + H_{\nu \nu}={{1}\over{2}}  \pmatrix{
-\Delta \cos 2\theta + A + B & \Delta\sin 2\theta + B_{\rm off} \cr
  \Delta\sin 2\theta + B^{\ast}_{\rm off}& \Delta \cos 2\theta - A - B \cr
},
\end{equation}
where $\Delta \equiv \delta m^2/(2E_\nu )$, with $E_{\nu}$ the
neutrino energy and $\delta m^2$ the appropriate vacuum neutrino 
mass-squared difference,
$A$ is the appropriate electron/nucleon background 
contribution, and $B$ and $B_{\rm off}$ are
the diagonal and off-diagonal 
contributions, respectively, from the neutrino background. 
(The terms $B_{\rm off}$ and $B_{\rm off}^*$ are complex conjugates of
each other and vanish in the case of active-sterile neutrino
transformation.) Note that a mass-level 
crossing or resonance
will occur if the following condition is satisfied:
\begin{equation}
\label{eqn:res}
\Delta\cos 2\theta = A+B.
\end{equation}
Table I gives 
appropriate expressions for the weak potentials
$A$ and $B$ for all possible cases of
$2\times 2$ active-sterile and active-active 
neutrino mixings in terms
of density $\rho$ (in ${\rm g}\,{\rm cm}^{-3}$), Avogadro's number 
$N_A$, $Y_e$, and the
effective neutrino numbers for each of the three active species 
relative to baryons, 
$Y_{\nu_e}$,
$Y_{\nu_\mu}$, $Y_{\nu_\tau}$. 

The effective neutrino number for 
species $\alpha$ ($\alpha=e,\ \mu,\ \tau$)
relative to baryons is $Y_{\nu_\alpha} =  (n_{\nu_\alpha}^{\rm eff} -
n_{\bar\nu_\alpha}^{\rm eff})/{\rho N_A}$, where for example, 
the effective number
density of neutrino species $\alpha$ at position $r$ encountered by a 
\lq\lq test\rq\rq\
neutrino traveling in a pencil of directions $\hat{\Omega}$ is
\cite{qfwmw93,fuller92,qianfor},
\begin{equation}
\label{eqn:eff}
n_{\nu_\alpha}^{\rm eff} \approx \sum_{\beta}
{{{1}\over{c \pi R^2_{\nu}}}  {{L_{\nu_\beta}}\over{\langle 
E_{\nu_\beta}\rangle}}
\int_{0}^{\infty}{dE_{\nu_\beta}}\int{{{d\Omega^\prime}\over{4\pi}}} 
f_{\nu_\beta}(
E_{\nu_\beta}) P_{\nu_\beta \rightarrow 
\nu_\alpha}(E_{\nu_\beta}, r,
\hat\Omega^\prime)
  (1-\hat{\Omega^\prime}\cdot\hat{\Omega})}.
\end{equation}
Here $\beta$ runs over all neutrino species to be
considered (including species $\alpha$), $f_{\nu_\beta}( 
E_{\nu_\beta})$ is
the {\it normalized} energy distribution function for neutrino 
species $\beta$ at its
birth position (the neutrino sphere of radius $R_\nu$), and 
$P_{\nu_\beta \rightarrow
\nu_\alpha}(E_{\nu_\beta}, r, \hat\Omega^\prime)$ is the 
probability for an initial $\nu_\beta$ to appear as a $\nu_\alpha$
when encountering the test neutrino at position $r$. 
In general, this
transformation probability will depend on the energy $E_{\nu_\beta}$ 
and direction
$\hat\Omega^\prime$ of the background neutrino $\nu_\beta$, as well as on the 
position $r$. The last
factor in Eq.\ (\ref{eqn:eff}) contains additional dependence on
the direction $\hat\Omega^\prime$ of the background neutrino.

For efficient conversion of neutrino flavors/types, evolution of the neutrino
amplitudes through the mass-level crossing (resonance) 
must be adiabatic. In turn,
this requires that the width of the resonance region
be large compared with the local
neutrino oscillation length.
The width of the resonance region is
\begin{equation}
\label{eqn:resw}
\delta r = {\cal{H}}
\tan{2\theta},
\end{equation}
where ${\cal{H}}$ is the scale height of the weak potential at resonance:
\begin{equation}
\label{eqn:scale}
{\cal{H}} \approx {\Bigg\vert {d\ln(A+B)\over dr}
\Bigg\vert}^{-1}_{\rm res}. 
\end{equation}
The adiabaticity parameter
$\gamma$ characterizing the evolution of the neutrino amplitudes
through resonance is
proportional to the ratio of the resonance width to the local
oscillation length:
\begin{equation}
\label{eqn:ad}
\gamma ={\delta m^2 \over{2E_\nu}}
{{\sin^22\theta}\over{\cos 2\theta}}{\cal{H}}\approx 4.6 {\left( {{\delta
m^2}\over{ 6\,{\rm
eV}^2}} \right)} {\left( {{25\,{\rm MeV}}\over{ E_\nu}} \right)}
\left({{\sin^22\theta}\over{{10}^{-3}}}\right)
{\left( {{\cal{H}}
\over{ 7.5\,{\rm km}}} \right)}.
\end{equation}
Note that $\gamma > 3$ corresponds to better than 99\% conversion of
neutrino flavors/types.

\subsection{The Case of Active-Sterile plus Active-Active Neutrino
Transformation}

In the context of the general scheme in Fig. 1 for neutrino masses
and mixings we make several specifications to facilitate our goal of 
removing the bulk of
the $\nu_e$ flux in the appropriate region 
above the neutrino sphere. First 
we invoke maximal
mixing between $\nu_\mu$ and $\nu_\tau$, which is 
consistent with the
SuperK atmospheric neutrino data. 
Furthermore, we assume that the vacuum mixing between
$\nu_{\mu,\tau}$ and $\nu_{e,s}$ is small.
In particular, we consider the following mixing scheme between
the flavor/type eigenstates and the vacuum mass eigenstates:
\begin{equation}
\label{eqn:matrix}
\pmatrix{|\nu_e\rangle\cr |\nu_s\rangle\cr 
        |\nu_\mu\rangle\cr |\nu_\tau\rangle\cr}=
\pmatrix{\cos\phi &\sin\phi\cos\omega &\sin\phi\sin\omega &0\cr
       -\sin\phi &\cos\phi\cos\omega &\cos\phi\sin\omega &0\cr
       0 &-{\sin\omega/\sqrt{2}} &{\cos\omega/\sqrt{2}} 
					&{1/\sqrt{2}}\cr
       0 &{\sin\omega/\sqrt{2}} &-{\cos\omega/\sqrt{2}}
                                        &{1/\sqrt{2}}\cr}
\pmatrix{|\nu_1\rangle\cr |\nu_2\rangle\cr 
	 |\nu_3\rangle\cr |\nu_4\rangle\cr}.
\end{equation}
In Eq. (\ref{eqn:matrix}), the vacuum mixing between $\nu_e$ and $\nu_s$
is largely governed by the angle $\phi$, while that between
$\nu_\mu$ and $\nu_\tau$ is chosen to be maximal by our assumption.
The angle $\omega$ essentially specifies the vacuum mixing between
$\nu_{\mu,\tau}$ and $\nu_{e,s}$.
For small vacuum mixing between these two doublets,
we require $\omega\ll 1$. In Eq. (\ref{eqn:matrix}),
all the CP-violating phases are ignored.

With the definition of
\begin{equation}
\label{eqn:numu}
|\nu_\mu^*\rangle\equiv{|\nu_\mu\rangle-|\nu_\tau\rangle\over\sqrt{2}},
\end{equation}
and
\begin{equation}
\label{eqn:nutau}
|\nu_\tau^*\rangle\equiv{|\nu_\mu\rangle+|\nu_\tau\rangle\over\sqrt{2}},
\end{equation}
Eq. (\ref{eqn:matrix}) can be rewritten as
\begin{equation}
\label{eqn:matrix2}
\pmatrix{|\nu_e\rangle\cr |\nu_s\rangle\cr
         |\nu_\mu^*\rangle\cr |\nu_\tau^*\rangle\cr}=
\pmatrix{\cos\phi &\sin\phi\cos\omega &\sin\phi\sin\omega &0\cr
       -\sin\phi &\cos\phi\cos\omega &\cos\phi\sin\omega &0\cr
       0 &-\sin\omega &\cos\omega &0\cr
       0 &0 &0 &1\cr}
\pmatrix{|\nu_1\rangle\cr |\nu_2\rangle\cr
        |\nu_3\rangle\cr |\nu_4\rangle\cr}.
\end{equation}
It is clear from Eq. (\ref{eqn:matrix2}) that $|\nu_\tau^*\rangle$ is a
mass eigenstate in vacuum, i.e., $H_{\rm v}|\nu_\tau^*\rangle\propto
|\nu_\tau^*\rangle$.
In fact, $|\nu_\tau^*\rangle$ is also an effective mass eigenstate
in the presence of electron/nucleon and neutrino backgrounds, i.e.,
$(H_{\rm v}+H_e+H_{\nu\nu})|\nu_\tau^*\rangle\propto|\nu_\tau^*\rangle$. 
This is
because the amplitude of forward scattering on electrons and nucleons is
the same for $\nu_\mu$ and $\nu_\tau$ while the effective number densities
of $\nu_\mu$ and $\nu_\tau$ for neutrino-neutrino scattering
above the neutrino sphere are the same
due to the symmetry in transformations concerning
$\nu_\mu$ and $\nu_\tau$.
Therefore, the evolution of $|\nu_\tau^\ast\rangle$ is decoupled from that
of $|\nu_s\rangle$, $|\nu_e\rangle$, and $\vert\nu_\mu^\ast\rangle$.

As discussed above, maximal mixing between $\nu_\mu$ and
$\nu_\tau$ allows us to reduce the problem of $4\times 4$ neutrino
mixing into that of mixing among $\nu_s$, $\nu_e$, and $\nu_\mu^*$.
In the region above, yet not too far away from
the neutrino sphere,
this $3\times3$ neutrino mixing problem is characterized by
the mass-level crossings for 
$\nu_\mu^*\rightleftharpoons\nu_s$
and $\nu_\mu^*\rightleftharpoons \nu_e$
transformations. Note that
although $\delta m_{\mu^*s}^2\approx\delta m_{\mu^*e}^2$
in our neutrino mass scheme shown in Fig. 1,
the weak potentials $A$ and $B$ in the neutrino propagation
Hamiltonian are quite different for 
$\nu_\mu^*\rightleftharpoons\nu_s$ and
$\nu_\mu^*\rightleftharpoons\nu_e$ transformations. The effect of
the weak potential $B$ from the neutrino background will be addressed in
Sec. IV. For simplicity, we will neglect $B$ in the following discussion.
According to Eq. (\ref{eqn:res}) and Table I, 
the mass-level crossing (resonance)
for $\nu_\mu^*\rightleftharpoons\nu_s$ transformation occurs when
\begin{equation}
\label{eqn:resmus}
{\delta m_{\mu^*s}^2\over 2E_\nu}\cos2\theta_{\mu^*s}=\sqrt{2}G_F\rho N_A
{1-Y_e\over 2},
\end{equation}
while that for $\nu_\mu^*\rightleftharpoons\nu_e$ transformation occurs
when
\begin{equation}
\label{eqn:resmue}
{\delta m_{\mu^*e}^2\over 2E_\nu}\cos2\theta_{\mu^*e}=\sqrt{2}G_F\rho N_A
Y_e.
\end{equation}
In Eqs. (\ref{eqn:resmus}) and (\ref{eqn:resmue}), $G_F$ is the Fermi
constant, and $\theta_{\mu^*s}$ and
$\theta_{\mu^*e}$ are the appropriate two-neutrino vacuum mixing angles.
From Eq. (\ref{eqn:matrix2}) we have $\theta_{\mu^*s}
\sim\omega\cos\phi\ll 1$ and $\theta_{\mu^*e}\sim\omega\sin\phi\ll 1$.

In our proposed scheme to enable
the $r$-process in neutrino-heated supernova ejecta by
disabling the alpha effect, we require that 
the $\nu_\mu^*
\rightleftharpoons \nu_s$ conversion of relatively high energy 
neutrinos take place well below
the weak freeze-out radius. The temperature at which weak 
freeze-out in the ejecta
occurs is $T_9^{\rm WFO} \approx 10$.
According to Eqs. (\ref{eqn:dens}) and (\ref{eqn:resmus}),
conversion below (at smaller radii than) the weak freeze-out 
radius then requires
\begin{equation}
\label{eqn:mus}
\delta m^2_{\mu^*s}
{\ \lower-1.2pt\vbox{\hbox{\rlap{$>$}\lower5pt\vbox{\hbox{$\sim$}}}}\ 
} 3.8\,{\rm eV}^2
{\left({{T_9^{\rm WFO}}\over{10}}\right)}^3
{\left({{E_\nu}\over{25\,{\rm MeV}}}\right)} {\left[{
(1-Y_e)/2\over{0.3}}\right]} 
S_{100}^{-1},
\end{equation}
where we have taken $g_s=11/2$, consistent with the conditions in the ejecta
for $T_9{\ 
\lower-1.2pt\vbox{\hbox{\rlap{$>$}\lower5pt\vbox{\hbox{$\sim$}}}}\ }
T_9^{\rm WFO}$,
and scaled the result assuming a typical value of
$Y_e = 0.4$ as obtained in
numerical supernova
models in the absence of neutrino transformation. 
Similarly, for the $\nu_\mu^*\rightleftharpoons \nu_e$ resonance to occur
below the weak freeze-out radius, we require
\begin{equation}
\label{eqn:mue}
\delta m^2_{\mu^*e}
{\ \lower-1.2pt\vbox{\hbox{\rlap{$>$}\lower5pt\vbox{\hbox{$\sim$}}}}\ 
} 5.1\,{\rm eV}^2
{\left({{T_9^{\rm WFO}}\over{10}}\right)}^3
{\left({{E_\nu}\over{25\,{\rm MeV}}}\right)}
{\left({Y_e\over{0.4 
}}\right)}
S_{100}^{-1}.
\end{equation}
Therefore, a value of 
$\delta m_{\mu^*s}^2\approx\delta m_{\mu^*e}^2\approx 6$
eV$^2$ consistent with the LSND data would fulfill the requirements in
Eqs. (\ref{eqn:mus}) and (\ref{eqn:mue}).

We note that for $Y_e > 1/3$,
the $\nu_\mu^*\rightleftharpoons \nu_s$ resonance will occur 
before (at higher density and temperature than)
the $\nu_\mu^*\rightleftharpoons \nu_e$ resonance
for a given neutrino
energy [cf. Eqs. (\ref{eqn:resmus}) and (\ref{eqn:resmue})]. 
Specifically, the temperature at resonance is
\begin{equation}
\label{eqn:tmus}
T_{9,\mu^*s}\approx 11.6
{\left( {{\delta m^2_{\mu^*s}}\over{6\ {\rm eV}^2}}
\right)}^{1/3} {\left( {25\ {\rm MeV}\over{E_\nu}}
\right)}^{1/3}
{\left[{0.3\over (1-Y_e)/2}\right]}^{1/3}
S_{100}^{1/3}
\end{equation}
for the $\nu_\mu^*\rightleftharpoons \nu_s$ case, and
\begin{equation}
\label{eqn:tmue}
T_{9,\mu^*e}\approx 10.6
{\left( {{\delta m^2_{\mu^*e}}\over{6\ {\rm eV}^2}}
\right)}^{1/3} {\left( {25\ {\rm MeV}\over{E_\nu}}
\right)}^{1/3}
{\left({0.4\over Y_e}\right)}^{1/3}
S_{100}^{1/3}
\end{equation}
for the $\nu_\mu^*\rightleftharpoons \nu_e$ case.
According to Eq. (\ref{eqn:wind1}), the radial separation between the
two resonances has a typical value of 2 km. As shown below, this is
much larger than the widths of both resonance regions.

The scale height of the weak potential for both
$\nu_\mu^*\rightleftharpoons \nu_s$ and
$\nu_\mu^*\rightleftharpoons \nu_e$ resonances can be approximated as
\begin{equation}
\label{eqn:scht}
{\cal{H}}\approx\left|{d\ln\rho\over dr}\right|^{-1}_{\rm res}
\approx {r_{\rm res}\over 3},
\end{equation}
where $r_{\rm res}$ is the radius for the relevant resonance, and we
have neglected the change in $Y_e$ compared with that in $\rho$.
To ensure adiabatic conversion and accommodate the LSND data at the
same time, we require $\sin^22\theta\sim 10^{-3}$ for
$\theta=\theta_{\mu^*s},\ \theta_{\mu^*e}$. [This is easily achieved
by having $\omega\sim 10^{-2}$ and $\phi\sim\pi/4$ in 
Eq. (\ref{eqn:matrix}).] From Eq. (\ref{eqn:resw}), the widths of the
$\nu_\mu^*\rightleftharpoons \nu_s$ and
$\nu_\mu^*\rightleftharpoons \nu_e$ resonance regions are
$(\delta r)_{\mu^*s}\sim (\delta r)_{\mu^*e}\sim 0.2$~km, much
smaller than the separation between the two resonances. Therefore,
we can treat the evolution of $\nu_e$ as unaffected by the
$\nu_\mu^*\rightleftharpoons \nu_s$ resonance and that of $\nu_s$
unaffected by the $\nu_\mu^*\rightleftharpoons \nu_e$ resonance.

So far we have restricted our discussion to the case of $Y_e>1/3$.
As the electron fraction $Y_e$ gets close to 1/3,
the $\nu_\mu^*\rightleftharpoons \nu_s$ and
$\nu_\mu^*\rightleftharpoons \nu_e$
resonances will approach each other and overlap. 
In addition, the weak potential governing the evolution of
neutrino amplitudes 
will be dominated
by the neutrino background for $Y_e\approx 1/3$. However,
as discussed in Sec. IV, the transformation
of neutrino flavors/types, coupled with the
expected rapid
expansion of the neutrino-heated ejecta, will
cause the actual value of $Y_e$ at a radius to differ significantly
from the equilibrium value corresponding to the local
$\nu_e$ and $\bar\nu_e$ fluxes.
For interesting ranges of $\delta m_{\mu^*s}^2$ and
$\delta m_{\mu^*e}^2$, the problematic neutrino evolution near 
$Y_e =1/3$ occurs well
beyond the weak freeze-out radius and does not affect our treatment
of the $\nu_\mu^*\rightleftharpoons \nu_s$ and
$\nu_\mu^*\rightleftharpoons \nu_e$ transformations discussed
previously.

We have also ignored the possibility of a $\nu_e 
\rightleftharpoons \nu_s$ resonance in the treatment of
active-sterile neutrino transformation.
This is because were this resonance to occur below the weak freeze-out
radius, neutrino evolution through 
this resonance
would be grossly non-adiabatic 
for the neutrino mass-squared difference ($\delta m^2_{es} {\
\lower-1.2pt\vbox{\hbox{\rlap{$<$}\lower5pt\vbox{\hbox{$\sim$}}}}\ } 
{10}^{-5}\,{\rm
eV}^2$) adopted to solve the solar neutrino problem in our neutrino
mixing scheme [see Eq. (\ref{eqn:ad})].

To summarize the discussion in this subsection,
we depict the squares of the effective neutrino 
masses as functions of
density in Fig. 2. Here we assume that 
$Y_e > 1/3$ and the order of the resonances
is as shown. As $Y_e$ decreases, the effective mass track for 
$\nu_\mu^*$
will steepen, while that for $\nu_e$ will flatten out. 
For $Y_e < 1/3$ this
trend will be so extreme that the order of the resonances will reverse.
As noted above and discussed in Sec. IV, 
for our adopted neutrino mixing parameters
this reversal does not occur below the weak freeze-out
radius in a typical neutrino-heated outflow.

\subsection{Removal of the $\nu_e$ Flux and Evasion of the Alpha Effect}

Based on the discussion in Sec. IIIB, we can regard the 
the $\nu_\mu^*\rightleftharpoons \nu_s$ and
$\nu_\mu^*\rightleftharpoons \nu_e$ resonances as filters of neutrino
flavors/types when neutrino evolution through these resonances is
adiabatic. Consider the evolution of an initial neutrino state
\begin{equation}
\label{eqn:nut0}
|\nu(t=t_0)\rangle=a_e|\nu_e\rangle+a_s|\nu_s\rangle
+a_{\mu^*}|\nu_\mu^*\rangle+a_{\tau^*}|\nu_\tau^*\rangle.
\end{equation}
At $t>t_{\mu^*s}$, i.e., after adiabatic propagation through the
$\nu_\mu^*\rightleftharpoons \nu_s$ resonance, the neutrino state
becomes
\begin{equation}
\label{eqn:nuts}
|\nu(t>t_{\mu^*s})\rangle\approx a_e'|\nu_e\rangle+a_s'|\nu_\mu^*\rangle
+a_{\mu^*}'|\nu_s\rangle+a_{\tau^*}'|\nu_\tau^*\rangle,
\end{equation}
where the primed coefficients differ from the corresponding unprimed
ones in Eq. (\ref{eqn:nut0}) only by a phase. Symbolically, the function
of the $\nu_\mu^*\rightleftharpoons \nu_s$ resonance can be described as
$|\nu_e\rangle\to|\nu_e\rangle$, $|\nu_s\rangle\to|\nu_\mu^*\rangle$,
$|\nu_\mu^*\rangle\to|\nu_s\rangle$, and 
$|\nu_\tau^*\rangle\to|\nu_\tau^*\rangle$.
At $t>t_{\mu^*e}>t_{\mu^*s}$, i.e., after adiabatic propagation through the
$\nu_\mu^*\rightleftharpoons \nu_e$ resonance, the neutrino state
becomes
\begin{equation}
\label{eqn:nute}
|\nu(t>t_{\mu^*e})\rangle\approx a_e''|\nu_\mu^*\rangle+a_s''|\nu_e\rangle
+a_{\mu^*}''|\nu_s\rangle+a_{\tau^*}''|\nu_\tau^*\rangle,
\end{equation}
where the double-primed coefficients differ from the corresponding unprimed
ones in Eq. (\ref{eqn:nut0}) again only by a phase. 
Symbolically, the function
of the $\nu_\mu^*\rightleftharpoons \nu_e$ resonance can be described as
$|\nu_e\rangle\to|\nu_\mu^*\rangle$, $|\nu_s\rangle\to|\nu_s\rangle$,
$|\nu_\mu^*\rangle\to|\nu_e\rangle$, and
$|\nu_\tau^*\rangle\to|\nu_\tau^*\rangle$.

The probabilities for the initial neutrino state in Eq. (\ref{eqn:nut0})
to evolve into the $|\nu_e\rangle$, $|\nu_s\rangle$, $|\nu_\mu^*\rangle$,
and $|\nu_\tau^*\rangle$ states after adiabatic propagation through the
$\nu_\mu^*\rightleftharpoons \nu_s$ and 
$\nu_\mu^*\rightleftharpoons \nu_e$ resonances can be obtained from
Eq. (\ref{eqn:nute}) as:
\begin{equation}
\label{eqn:pnue}
P_{\nu_e}(t>t_{\mu^*e})\approx\Big|\langle\nu_s|\nu(t=t_0)\rangle\Big|^2,
\end{equation}
\begin{equation}
\label{eqn:pnus}
P_{\nu_s}(t>t_{\mu^*e})\approx
\Big|\langle\nu_\mu^*|\nu(t=t_0)\rangle\Big|^2,
\end{equation}
\begin{eqnarray}
\label{eqn:pnumu}
P_{\nu_\mu}(t>t_{\mu^*e})&\approx&
\Big|\langle\nu_\mu|\nu_\mu^*\rangle\langle\nu_e|\nu(t=t_0)\rangle+
\langle\nu_\mu|\nu_\tau^*\rangle\langle\nu_\tau^*|\nu(t=t_0)\rangle
\exp(i\Phi)\Big|^2\cr
& & \cr
&\approx&{1\over 2}\left(\Big|\langle\nu_e|\nu(t=t_0)\rangle\Big|^2
+\Big|\langle\nu_\tau^*|\nu(t=t_0)\rangle\Big|^2\right),
\end{eqnarray}
and
\begin{eqnarray}
\label{eqn:pnutau}
P_{\nu_\tau}(t>t_{\mu^*e})&\approx&
\Big|\langle\nu_\tau|\nu_\mu^*\rangle\langle\nu_e|\nu(t=t_0)\rangle+
\langle\nu_\tau|\nu_\tau^*\rangle\langle\nu_\tau^*|\nu(t=t_0)\rangle
\exp(i\Phi)\Big|^2\cr
& & \cr
&\approx&{1\over 2}\left(\Big|\langle\nu_e|\nu(t=t_0)\rangle\Big|^2
+\Big|\langle\nu_\tau^*|\nu(t=t_0)\rangle\Big|^2\right).
\end{eqnarray}
In Eqs. (\ref{eqn:pnumu}) and (\ref{eqn:pnutau}) the second 
approximation is obtained by averaging over the phase $\Phi$.

From Eqs. (\ref{eqn:pnue})--(\ref{eqn:pnutau}) we see that
after adiabatic evolution through 
the $\nu_\mu^*\rightleftharpoons \nu_s$ and
$\nu_\mu^*\rightleftharpoons \nu_e$ resonances,
(1) a $\nu_e$ emitted from the neutron star 
[i.e., $|\nu(t=t_0)\rangle=|\nu_e\rangle$] would become 50\% $\nu_\mu$
and 50\% $\nu_\tau$;
(2) a $\nu_\mu$ emitted from the neutron star 
[i.e., $|\nu(t=t_0)\rangle=|\nu_\mu\rangle$] would become 50\% $\nu_s$,
25\% $\nu_\mu$, and 25\% $\nu_\tau$; and
(3) a $\nu_\tau$ emitted from the neutron star
[i.e., $|\nu(t=t_0)\rangle=|\nu_\tau\rangle$] would become 50\% $\nu_s$,
25\% $\nu_\mu$, and 25\% $\nu_\tau$.
Clearly, as no sterile neutrinos are emitted from the neutron star,
the $\nu_e$ flux would be removed at those energies for which
adiabatic evolution through 
the $\nu_\mu^*\rightleftharpoons \nu_s$ and
$\nu_\mu^*\rightleftharpoons \nu_e$ resonances is completed.
Therefore, if such neutrino evolution can be engineered to 
occur below the weak
freeze-out radius and convert most of the $\nu_e$ emitted from 
the neutron star to other 
species before the
electron fraction falls near or below 1/3, then we will have 
succeeded in disabling
the alpha effect.

\section{The Evolution of the Electron Fraction with
Active-Sterile plus Active-Active Neutrino Transformation}

Consider the $Y_e$ evolution of a mass/fluid element
moving outward from the neutron star surface 
(i.e., the neutrino sphere). 
In the absence of neutrino flavor/type transformation, this fluid
element would be
irradiated by the $\nu_e$, $\bar\nu_e$, $\nu_\mu$, $\bar\nu_\mu$,
$\nu_\tau$, and $\bar\nu_\tau$ fluxes emitted from
the neutrino sphere throughout its progress toward ejection. (There
should not be an {\it appreciable} flux of sterile neutrinos coming 
from the neutron star interior.
Though a small sterile neutrino flux will not affect our 
conclusions, it is nonetheless
interesting to note that for our assumed scheme of neutrino masses
and mixings, 
matter effects inside the proto-neutron star
should suppress
strongly the production of sterile neutrinos.) Of course,
the occurrence of the 
$\nu_\mu^*
\rightleftharpoons \nu_e$ resonance will cause the fluid 
element to
experience deviation of its $Y_e$ evolution from the standard case 
with no neutrino flavor/type
transformation.

As discussed in Sec. IIIB, the $\nu_\mu^*\rightleftharpoons\nu_s$
resonance occurs before (at higher temperature than) the
$\nu_\mu^*\rightleftharpoons\nu_e$ resonance for $Y_e>1/3$.
Further, neutrinos with lower energies generally go through the two
resonances before those with higher energies. Consequently,
the two resonances will sweep through the relevant
neutrino energy distribution functions when viewed
in the frame of a
fluid element as it moves from the neutrino sphere at 
high temperature towards regions of lower temperature.
The energy-position of the
$\nu_\mu^*
\rightleftharpoons
\nu_e$ resonance within the distribution functions, 
i.e., the resonance 
energy $E_\nu^{\rm
RES}$, is related to temperature $T_9$ and radius
$r_6$ through the resonance condition as
\begin{eqnarray}
\label{eqn:thresh}
E_\nu^{\rm RES} &\approx&
29.4\,{\rm MeV}{\left( {{\delta m^2_{\mu^*
e}}\over{6\,{\rm eV}^2}}
\right)} {\left[{0.4\over Y_e+Y_{\nu_e}
-(Y_{\nu_\mu}+Y_{\nu_\tau})/2}\right]}
\left({10\over T_9}\right)^3S_{100}\cr
& & \cr
&\approx&
2.58\,{\rm MeV}{\left( {{\delta m^2_{\mu^* 
e}}\over{6\,{\rm eV}^2}}
\right)} {\left[{0.4\over Y_e+Y_{\nu_e}
-(Y_{\nu_\mu}+Y_{\nu_\tau})/2}\right]}
{\left({1.4\,M_\odot\over M_{\rm NS}}\right)^3}
S_{100}^4 r_6^3.
\end{eqnarray}

When a fluid element reaches a given radius,
we can regard all $\nu_e$ emitted from the neutrino sphere
with energies less than the corresponding
$E_\nu^{\rm RES}$ at this radius as having changed to either $\nu_\mu$ 
or $\nu_\tau$. Note that no conversion into $\nu_e$ occurs as
the $|\nu_\mu^*\rangle$ component of the
$\nu_\mu$ or $\nu_\tau$ emitted from the neutrino sphere 
with the relevant energies 
will have been converted into sterile
neutrinos before reaching the $\nu_\mu^*
\rightleftharpoons
\nu_e$ resonance. 
(Adiabatic neutrino 
evolution
through the $\nu_\mu^*
\rightleftharpoons
\nu_s$ and $\nu_\mu^*
\rightleftharpoons
\nu_e$ resonances is assumed here.) 
The $\nu_e$ and $\bar\nu_e$ capture rates in the fluid element, 
$\lambda_{\nu_e n}$ and 
$\lambda_{\bar\nu_e p}$, respectively, will both decrease
with increasing radius due to the geometric dilution of the 
neutrino fluxes. However,
as $E_\nu^{\rm RES}$ increases, conversion into $\nu_\mu$ or $\nu_\tau$
will further decrease the $\nu_e$ flux in addition to the geometric
dilution. Consequently, the rate $\lambda_{\nu_e n}$ for raising $Y_e$
will decrease more with increasing radius than
the rate $\lambda_{\bar\nu_e p}$ for lowering $Y_e$.
One would then expect
$Y_e$ to be lowered progressively as the fluid element moves towards
larger radius. 
If $Y_e {\
\lower-1.2pt\vbox{\hbox{\rlap{$<$}\lower5pt\vbox{\hbox{$\sim$}}}}\ } 
1/3$ is reached, the order
of the $\nu_\mu^*\rightleftharpoons\nu_s$ and
$\nu_\mu^*\rightleftharpoons\nu_e$ resonances will reverse 
(see Sec. IIIB) and further
transformation of $\nu_e$ will cease. In principle this 
could be worrisome as the evasion of
the alpha effect requires that the bulk of the 
$\nu_e$ flux over a substantial range of energies
be transformed away prior to the point of alpha-particle formation.

However, in practice 
the defeat of the
alpha effect through removal of the $\nu_e$ flux will be achieved
for the expected conditions in the neutrino-heated ejecta and 
plausible neutrino mixing parameters. This is because $\nu_e$ 
conversion
coupled with the expected rapid expansion of the ejecta
will cause the actual value of $Y_e$ at a given radius to differ 
significantly from the equilibrium value corresponding to the local
$\nu_e$ and $\bar\nu_e$ fluxes. This equilibrium value 
at radius $r$ is defined as
\begin{equation}
\label{eqn:yeq}
Y_e^{\rm EQ}(r)\equiv{\lambda_{\nu_en}(r)\over
\lambda_{\nu_en}(r)+\lambda_{\bar\nu_ep}(r)}.
\end{equation}
As discussed above, in the presence of neutrino flavor/type
transformation, the decrease of $\lambda_{\nu_en}$ due to
conversion of $\nu_e$ into $\nu_\mu$ or $\nu_\tau$ in
addition to the geometric decrease common to both
$\lambda_{\nu_en}$ and $\lambda_{\bar\nu_ep}$ tends to
lower $Y_e^{\rm EQ}$ at larger radius. Thus we have
$dY_e^{\rm EQ}/dr<0$.

On the other hand, the $Y_e$ evolution of a fluid element 
prior to the point of
alpha-particle formation is governed by
\begin{equation}
\label{eqn:yev}
v(r){dY_e(r)\over dr}=\lambda_{\nu_en}(r)-
[\lambda_{\nu_en}(r)+\lambda_{\bar\nu_ep}(r)]Y_e(r),
\end{equation}
where $v(r)$ is the outflow velocity at radius $r$.
To first order, the solution to Eq. (\ref{eqn:yev}) is
\begin{equation}
\label{eqn:ye1}
Y_e(r)\approx Y_e^{\rm EQ}(r)-
{v(r)\over\lambda_{\nu_en}(r)+\lambda_{\bar\nu_ep}(r)}
{dY_e^{\rm EQ}(r)\over dr}.
\end{equation}
As the fluid element moves to larger radius, 
the conversion ``front'' at
$E_\nu^{\rm RES}$ moves towards the high-energy end of the $\nu_e$
energy distribution function (shown schematically in
Figure 3), causing $Y_e^{\rm EQ}$ to decrease. 
However, according to Eq. (\ref{eqn:ye1}), this
acts to increase $Y_e$ above
$Y_e^{\rm EQ}$. Further, a larger increase is obtained for
a larger outflow velocity $v$ corresponding to a more rapid
expansion of the fluid element.
Consequently, flavor evolution of
$\nu_e$ with energies important for determining $Y_e$
can occur with $Y_e {\
\lower-1.2pt\vbox{\hbox{\rlap{$>$}\lower5pt\vbox{\hbox{$\sim$}}}}\ }
1/3$ in the rapidly expanding neutrino-heated ejecta
for a range of outflow conditions and neutrino mixing parameters.

As indicated in Eq. (\ref{eqn:thresh}), 
the resonance energy $E_\nu^{\rm RES}$
at a given radius depends on the effective neutrino numbers relative to
baryons $Y_{\nu_e}$, $Y_{\nu_\mu}$, and $Y_{\nu_\tau}$. These quantities
determine the effects of the neutrino backgrounds on neutrino flavor/type
evolution. Using Eq. (\ref{eqn:eff}) 
and assuming adiabatic neutrino evolution,
we find that for a radially travelling neutrino, 
\begin{equation}
\label{eqn:ynue}
Y_{\nu_e} \approx A_{\nu_e} \left[ L_{\nu_e,50}
\left({10\,{\rm MeV}}\over{\langle E_{\nu_e}\rangle}\right) 
\int_{E_\nu^{\rm
RES}}^{\infty}{f_{\nu_e}(E_{\nu_e})dE_{\nu_e}} 
- L_{\bar\nu_e,50} \left({10\,{\rm 
MeV}}\over{\langle
E_{\bar\nu_e}\rangle}\right)\right],
\end{equation}
where 
\begin{equation}
\label{eqn:A}
A_{\nu_e} =0.723 
{\left({{1.4\,{\rm M}_\odot}\over{M_{\rm NS}}}\right)}^3 
{S_{100}^4 
r_6^3\over R_{\nu,6}^2}
\left[ 1-\sqrt{
1-{R_{\nu,6}^2\over r_6^2}}\right]^2,
\end{equation}
and
\begin{equation}
\label{eqn:fd}
f_{\nu_e}(E_{\nu_e}) = {{1}\over{T_{\nu_e}^3 F_2(\eta_{\nu_e})}}
{{E_{\nu_e}^2 }\over{\exp{(E_{\nu_e}/T_{\nu_e} -\eta_{\nu_e})} +1}}
\end{equation}
is the normalized $\nu_e$ energy distribution function.
In Eqs. (\ref{eqn:ynue}) and (\ref{eqn:A})
$L_{\nu_e,50}$ and $L_{\bar\nu_e,50}$
are the neutrino energy luminosities in units of
${10}^{50}\,{\rm ergs}\,{\rm s}^{-1}$ and $R_{\nu,6}$ is the 
neutrino-sphere radius in units of
${10}^6\,{\rm cm}$. 
In Eq. (\ref{eqn:fd}) 
$F_2(\eta_{\nu_e})\equiv\int_0^\infty x^2/[\exp(x-\eta_{\nu_e})+1]dx$ 
is the second-order relativistic Fermi integral, and
the two parameters $T_{\nu_e}$ and
$\eta_{\nu_e}$ 
can be specified by fitting 
the first two energy moments of the numerical $\nu_e$
energy spectrum fom supernova neutrino transport calculations.
Expressions for $Y_{\nu_\mu}$ and $Y_{\nu_\tau}$ can be obtained
in similar manner.

For $r
{\ \lower-1.2pt\vbox{\hbox{\rlap{$>$}\lower5pt\vbox{\hbox{$\sim$}}}}\
}1.5R_\nu$, Eq. (\ref{eqn:A}) can be rewritten as
\begin{equation}
\label{eqn:AA}
A_{\nu_e}\approx 0.08\left({T_9\over 10}\right)
{\left({{1.4\,{\rm M}_\odot}\over{M_{\rm NS}}}\right)}^4
R_{\nu,6}^2S_{100}^5.
\end{equation}
Taking, for example,
$L_{\nu_e}\sim L_{\bar\nu_e}\sim 10^{50}$~erg~s$^{-1}$,
$\langle E_{\nu_e}\rangle\approx 11$~MeV, and
$\langle E_{\bar\nu_e}\rangle\approx 16$~MeV, which are
plausibly characteristic of neutrino emission at very late times,
we have $|Y_{\nu_e}|\sim 0.01$ at $T_9\sim 10$ for $R_{\nu,6}=1$
and $S_{100}=0.7$. Therefore, compared with typical values of
$Y_e\sim 0.4$, $Y_{\nu_e}$, $Y_{\nu_\mu}$, and $Y_{\nu_\tau}$
can be neglected. Further, the effects of the neutrino backgrounds on
neutrino flavor/type evolution are especially small for low
values of $S_{100}$ [see Eq. (\ref{eqn:AA})]. 
This is because important neutrino flavor/type evolution occurs
over a relatively narrow range of temperatures and for
a lower entropy, 
these temperatures correspond to
larger radii [see Eq. (\ref{eqn:wind1})]
where neutrino fluxes are smaller.

In order to disable the alpha effect, we would like to have the bulk of
the $\nu_e$ flux at $E_{\nu_e}<20$~MeV (see Fig. 3) transformed away
before the point of alpha-particle formation ($T_9\sim 10$).
This requires that the $\nu_e$ conversion front be at 
$E_\nu^{\rm RES}\approx 20$~MeV when the neutrino-heated ejecta 
has reached the radius corresponding to $T_9\approx 10$.
With a typical entropy of $S_{100}=0.7$ 
and values of $Y_e\sim 0.4\gg Y_{\nu_e},\,Y_{\nu_\mu},\,Y_{\nu_\tau}$ 
in the ejecta, we see from
Eq. (\ref{eqn:thresh}) that this requirement is indeed fulfilled
for $\delta m_{\mu^*e}^2\approx 6$~eV$^2$.

The typical evolution of $Y_e$ in the neutrino-heated ejecta with
$\nu_\mu^*\rightleftharpoons\nu_s$ plus
$\nu_\mu^*\rightleftharpoons\nu_e$ transformation
can then be summarized as follows.
At $T_9\sim 20$, the electron fraction 
has a typical value of $Y_e\sim 0.4$ in equilibrium
with the $\nu_e$ and $\bar\nu_e$ fluxes. As the ejecta moves to regions
of lower temperature, the $\nu_e$ conversion front at $E_\nu^{\rm RES}$
moves towards the high-energy end of the $\nu_e$ energy distribution
function and removes the $\nu_e$ flux at $E_{\nu_e}<E_\nu^{\rm RES}$.
This suppresses the destruction of neutrons and lowers $Y_e$ somewhat
below $\sim 0.4$. However, $\nu_e$ conversion 
coupled with the expansion of the ejecta causes rapid change in the 
instantaneous
equilibrium $Y_e$ value, $Y_e^{\rm EQ}$, and maintains
$Y_e>Y_e^{\rm EQ}$. Consequently, conversion of $\nu_e$ with energies
important for determining $Y_e$ is completed at $Y_e>1/3$. 
When the ejecta reaches the point of alpha-particle formation
at $T_9\sim 10$, the $\nu_e$ conversion front is at 
$E_\nu^{\rm RES}\approx 20$~MeV and the bulk of the $\nu_e$ flux
has been removed. This then defeats the alpha effect which would
drive $Y_e$ close to 0.5 if a significant flux of $\nu_e$ existed
to destroy neutrons at $T_9
{\ \lower-1.2pt\vbox{\hbox{\rlap{$<$}\lower5pt\vbox{\hbox{$\sim$}}}}\
}10$.

Therefore, our scenario of $\nu_\mu^*\rightleftharpoons\nu_s$ plus
$\nu_\mu^*\rightleftharpoons\nu_e$ transformation ensures that
$1/3<Y_e
{\ \lower-1.2pt\vbox{\hbox{\rlap{$<$}\lower5pt\vbox{\hbox{$\sim$}}}}\
}0.4$ can be obtained and {\it maintained} before the onset of 
rapid neutron
capture in the neutrino-heated ejecta. While this scenario
does not lead to very low values of $Y_e$, it nevertheless
guarantees a range of $Y_e$ that will lead to a successful $r$-process
when combined with the appropriate entropy $S$ and the dynamic expansion
timescale $\tau_{\rm DYN}$ in the ejecta.

\section{Conclusions}

The alpha effect is the single biggest impediment to
obtaining the necessary conditions for $r$-process nucleosynthesis in 
neutrino-heated ejecta
from supernovae. In fact, this effect may also be an obstacle for
$r$-process nucleosynthesis in neutrino-heated ejecta from 
neutron-star mergers. The 
central problem is
that the $\nu_e$ flux causing ejection of baryons from deep in the
gravitational potential well of the neutron star will 
destroy
neutrons via $\nu_e + n\rightarrow p+e^-$ in the regions of
alpha-particle formation. This destruction of neutrons then renders
the subsequent neutron capture process
incapable of
producing the heavy $r$-process elements.
We have suggested a scheme of neutrino 
masses and mixings
which could solve this conundrum by removing 
the $\nu_e$ flux through matter-enhanced
$\nu_{\mu,\tau}\rightleftharpoons\nu_s$ plus
$\nu_{\mu,\tau}\rightleftharpoons\nu_e$ transformation above
the regions of efficient neutrino heating but below the regions
of alpha-particle formation.
This scheme which rescues $r$-process
nucleosynthesis in neutrino-heated ejecta originally was not 
constructed for this
purpose. Rather, it was designed to explain simultaneously the solar 
neutrino data and the
anomalous atmospheric $\nu_\mu/\nu_e$ ratio and to allow for a hot 
component of dark matter.
Subsequently, it also explained the LSND signal. In fact, this 
scheme with a 
maximally-mixed $\nu_\mu$-$\nu_\tau$ doublet split from a lower-mass 
$\nu_e$-$\nu_s$
doublet by $(\delta m^2)_{\rm LSND}{\ 
\lower-1.2pt\vbox{\hbox{\rlap{$>$}\lower5pt\vbox{\hbox{$\sim$}}}}\ }
1\,{\rm eV}^2$ may be the only one which can escape elimination by 
recent interpretations
of the SuperK atmospheric neutrino data \cite{barg,BGG} and by Big Bang 
Nucleosynthesis considerations
\cite{SFA}.

While we do not know if there is a way other than invoking neutrino mixing
to circumvent
the alpha effect in neutrino-heated ejecta,
attempts at ``astrophysical'' dodge of this effect made so far 
(see, e.g., Ref. \cite{gr}) seem
to be finely tuned at best.
Absent a non-neutrino-physics escape from
the alpha effect, we could draw interesting inferences 
about Galactic chemical
evolution were future experiments to reveal a neutrino mass scheme 
{\it other}
than the one that aids the $r$-process. In that case, for example, we 
could be forced to
re-think the origin of the bulk of the $r$-process material in the 
Galaxy. In turn, this
could have consequences for our understanding of the rates and 
physics of, e.g.,
neutron-star mergers. Or we could be forced to re-think 
electromagnetic ejection of
material from supernovae.

In addition to the $r$-process connection, our scheme of neutrino masses
and mixings has interesting
consequences for detection of neutrinos from future Galactic supernovae.
As the solar neutrino problem is solved by 
$\nu_e \rightleftharpoons \nu_s$ mixing in this scheme, sterile neutrinos
produced by the $\nu_{\mu,\tau}\rightleftharpoons\nu_s$ transformation
near the neutron star will be converted into $\nu_e$. Therefore,
at late
times of the supernova process,
there may be a 
significant {\it increase}
in the average $\nu_e$ energy with no accompanying increase in the 
average $\bar\nu_e$ energy.
Furthermore, it has been shown that neutrino flavor transformation
has important effects on the dynamics of supernova explosion
\cite{fuller92,qianfor,mezza,nunokawa}. 
Conceivably, neutrino mixing also affects the
nucleosynthesis 
discussed in Ref. \cite{hoff},
which occurs shortly after the supernova explosion.
Currently, we are investigating the effects of our scheme of neutrino 
masses and mixings at these early times.

From this work, one conclusion specific to particle physics
is that evasion of the 
alpha effect and, hence, robust
production of the $r$-process elements in neutrino-heated ejecta seem to 
require at least one
light sterile neutrino species. We have shown here that 
one neutrino mass scheme to disable the alpha effect
has a maximally-mixed $\nu_\mu$-$\nu_\tau$ doublet split from 
a lower-mass
$\nu_e$-$\nu_s$ doublet. However, another scheme \cite{mfbf},
which has 
three light, nearly-degenerate
active neutrinos and a heavier sterile neutrino species,
has also been suggested to do the same. Though 
this latter scheme has
the attractive feature of a significant increase in the neutron 
excess,
considerations 
that do not concern the $r$-process
argue against it.  Finally, 
if we adopt the neutrino mass 
scheme of this paper,
then 
the splitting 
between the doublets
must be $\delta m^2
{\ \lower-1.2pt\vbox{\hbox{\rlap{$>$}\lower5pt\vbox{\hbox{$\sim$}}}}\
}1$~eV$^2$
in order to have the beneficial effects on the $r$-process.
Note that this splitting is within the LSND range 
and most likely in
the upper sector of it.  Of course, these conclusions presuppose that 
{\it some} of the
$r$-process material in the Galaxy originated in neutrino-heated 
ejecta (from either
supernovae or neutron-star mergers) and that there is no conventional 
astrophysical fix
for the alpha effect.

Our conclusions will be tested by neutrino experiments as well as
and astrophysical observations on $r$-process nucleosynthesis.  
In any case, it is both surprising 
and tantalizing that
mixings among active and sterile neutrino species with small masses
can have such
profound effects on the physics of astrophysical objects and the 
synthesis of the
heaviest elements.

\acknowledgements

We would like to acknowledge discussions with
A. B. Balantekin, J. Fetter, G. C. McLaughlin, M. Patel,
and J. R. Wilson. This work was
partially supported by DOE Grant DE-FG03-91ER40618 at
UCSB, by NSF Grant PHY98-00980 at UCSD,
and by DOE Grant DE-FG02-87ER40328 at UMN.

\pagebreak
\begin{figure}
\caption{The neutrino mass scheme discussed in this paper. A doublet of 
(near)-maximally-mixed
$\nu_\mu$ and $\nu_\tau$ neutrinos with mass-squared difference 
$\delta m_{\mu\tau}^2 \sim {10}^{-2}$~eV$^2$ is split from a doublet 
of lower-mass
$\nu_e$ and $\nu_s$ by a mass-squared difference 
$\delta m_{\rm doublets}^2$.}\noindent The
lower-mass doublet can be arranged (lower right) with 
maximal (or near maximal) 
$\nu_e\rightleftharpoons \nu_s$
vacuum mixing to give a ``just 
so'' solar neutrino
solution, which would require 
$\delta m^2_{es} \sim {10}^{-10}$~eV$^2$.
Alternatively (lower left), the sterile neutrino could be heavier than the
$\nu_e$ with $\delta m^2_{es}
{\ \lower-1.2pt\vbox{\hbox{\rlap{$<$}\lower5pt\vbox{\hbox{$\sim$}}}}\
}{10}^{-5}$~eV$^2$
to give a matter-enhanced (MSW) solution to the solar neutrino 
problem. We assume that
$\delta m_{\rm doublets}^2$ is much larger than the mass-squared 
splittings within the
doublets.
\end{figure}
\begin{figure}
\caption{Cartoon of the instantaneous neutrino mass levels (effective 
mass-squared $m^2_{\rm
eff}$) as functions of matter density
$\rho$ for 
$Y_e > 1/3$.}
\end{figure}
\begin{figure}
\caption{Example $\nu_e$ and $\nu_{\mu,\tau}$ energy distribution 
functions.
Here we take these functions to be of the form in
Eq. ({\protect\ref{eqn:fd}}) with $T_{\nu_e}=2.75$~MeV,
$\eta_{\nu_e}=3$ (corresponding to
$\langle
E_{\nu_e}\rangle = 11\,{\rm MeV}$) and
$T_{\nu_{\mu,\tau}}=6.76$~MeV, $\eta_{\nu_{\mu,\tau}}=3$
(corresponding to
$\langle E_{\nu_{\mu,\tau}}\rangle = 27\,{\rm MeV}$).
The resonance energy $E_{\nu}^{\rm RES}$ for neutrino flavor/type
transformation sweeps from low to high 
energy through the
neutrino energy 
distribution functions as a fluid element moves away from the 
neutron star.}
\end{figure}

\pagebreak

\begin{table}
\caption{Weak potentials derived from neutrino forward scattering
on electron/nucleon ($A$) and neutrino ($B$) backgrounds
for various channels of neutrino 
transformation.
The corresponding antineutrino transformation channels have 
opposite signs for
$A$ and $B$. Here $N_A$ is Avogadro's number,
$G_F$ is the Fermi constant, $\rho$ is the matter density,
$Y_e$ is
the electron fraction, and 
$Y_{\nu_e}$, $Y_{\nu_\mu}$, and $Y_{\nu_\tau}$
are the effective neutrino numbers for the corresponding species
relative to baryons.} 
\begin{center}
\begin{tabular}{c c c c c c c c c}
{Channel} &{$A$} &{$B$}
\\
\hline \\
$\nu_e \rightleftharpoons \nu_s$ & $(3\sqrt{2}/2)G_{F} 
\rho N_{A} \left(
Y_e -1/3\right)$ &
$\sqrt{2} G_{F} \rho N_{A} \left(
2 Y_{\nu_e} + Y_{\nu_\mu}+Y_{\nu_\tau} \right)$ \\
\hline \\
$\nu_\mu \rightleftharpoons \nu_s$ & $(\sqrt{2}/2)G_{F} 
\rho N_{A}
\left( Y_e -1\right)$ &
$\sqrt{2} G_{F} \rho N_{A} \left(
  Y_{\nu_e} + 2 Y_{\nu_\mu}+Y_{\nu_\tau} \right)$ \\
\hline \\
$\nu_\tau \rightleftharpoons \nu_s$ & $(\sqrt{2}/2)G_{F} 
\rho N_{A}
\left( Y_e -1\right)$ &
$\sqrt{2} G_{F} \rho N_{A} \left(
  Y_{\nu_e} +  Y_{\nu_\mu}+ 2 Y_{\nu_\tau} \right)$ \\
\hline \\
$\nu_\mu \rightleftharpoons\nu_e$ & $\sqrt{2} G_{F} \rho N_{ 
A} Y_e$ & $\sqrt{2}
G_{F} \rho N_{A} \left( Y_{\nu_e}-Y_{\nu_\mu}\right)$ \\
\hline \\
$\nu_\tau \rightleftharpoons\nu_e$ & $\sqrt{2} G_{F} \rho N_{ 
A} Y_e$ & $\sqrt{2}
G_{F} \rho N_{A} \left( Y_{\nu_e}-Y_{\nu_\tau}\right)$ \\
\hline \\
$\nu_\mu \rightleftharpoons\nu_\tau$ & 0 & $\sqrt{2}
G_{F} \rho N_{A} \left( Y_{\nu_\mu}-Y_{\nu_\tau}\right)$ \\
\end{tabular}
\end{center}
\end{table}

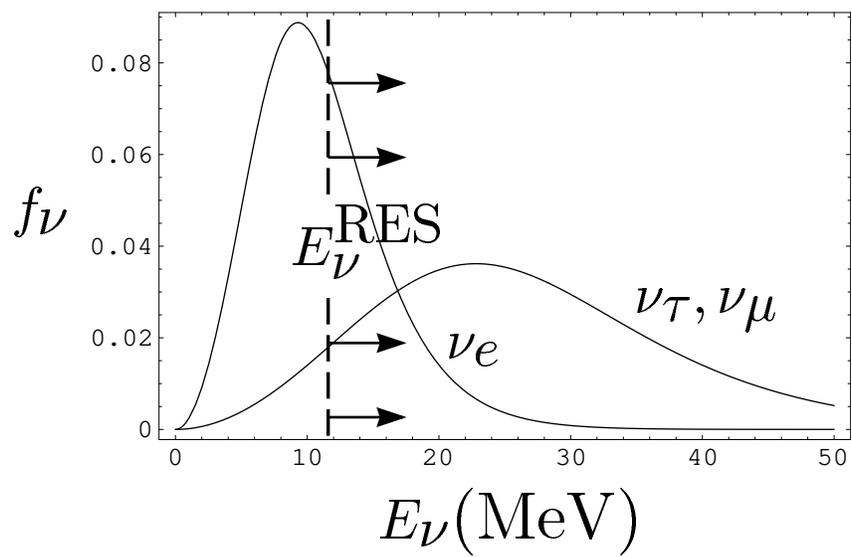
\end{document}